\documentclass[
 reprint,
 superscriptaddress,
 amsmath,amssymb,
 aps,
 floatfix,
]{revtex4-2}

\usepackage{xpatch}
\makeatletter
\xpatchcmd{\@ssect@ltx}{\@xsect}{\protected@edef\@currentlabelname{#8}\@xsect}{}{}
\xpatchcmd{\@sect@ltx}{\@xsect}{\protected@edef\@currentlabelname{#8}\@xsect}{}{}
\makeatother

\usepackage{graphicx}
\usepackage{dcolumn}
\usepackage{bm}
\usepackage{physics}
\usepackage{csquotes}
\usepackage{mathtools}
\usepackage{amsfonts}
\usepackage{amssymb}
\usepackage{amsmath}
\usepackage[caption=false,labelformat=empty]{subfig}
\usepackage{diagbox}
\usepackage{dsfont}
\usepackage{hyperref}
\usepackage{placeins}
\usepackage{xspace}
\usepackage{nameref}
\usepackage{orcidlink}

\usepackage{makecell}

\usepackage{nicematrix}
\NiceMatrixOptions{cell-space-limits = 2pt}
\hypersetup{linktocpage,colorlinks,citecolor={blue},pdfdisplaydoctitle=true,pdfpagemode=UseOutlines,bookmarksnumbered=true}
\usepackage{verbatim}

\newcommand*{\id}{\ensuremath{\mathds{1}}}

\newcommand{\mat}[1]{\mathbf{#1}}
\newcommand{\site}[1]{\ensuremath{\mathbf{#1}}}
\newcommand{\uvec}[1]{\ensuremath{\hat{\mathbf{#1}}}}

\newcommand{\aqa}{$\langle aQa ^L\rangle $ Applied Quantum Algorithms, Universiteit Leiden}
\newcommand{\lorentz}{Instituut-Lorentz, Universiteit Leiden, Niels Bohrweg 2, 2333 CA Leiden, Netherlands}
\newcommand{\ulm}{Institute for Complex Quantum Systems, Ulm University, 89069 Ulm, Germany}
\newcommand{\iqst}{Center for Integrated Quantum Science and Technology (IQST), Ulm-Stuttgart, Germany}
\newcommand{\racah}{Racah Institute of Physics, The Hebrew University of Jerusalem, Givat Ram, Jerusalem 91904, Israel}
\newcommand{\telaviv}{School of Physics and Astronomy, Tel Aviv University, Tel Aviv 6997801, Israel}

\begin{document}

\title{Projected Entangled Pair States for Lattice Gauge Theories with Dynamical Fermions}

\author{Ariel Kelman\,\orcidlink{0000-0002-7710-6538}}
\email{ariel.kelman@mail.huji.ac.il} 
\thanks{Corresponding author.}
\affiliation{\racah}
\author{Umberto Borla\,\orcidlink{0000-0002-4224-5335}}
\affiliation{\racah}
\affiliation{Max Planck Institute of Quantum Optics, 85748 Garching, Germany}
\affiliation{Munich Center for Quantum Science and Technology (MCQST), 80799 Munich, Germany}
\author{Patrick Emonts\,\orcidlink{0000-0002-7274-4071}}
\email{patrick.emonts@uni-ulm.de} 
\thanks{Corresponding author.}
\affiliation{\ulm}
\affiliation{\iqst}
\affiliation{\lorentz}
\affiliation{\aqa}
\author{Erez Zohar\,\orcidlink{0000-0001-6993-6569}}
\affiliation{\racah}
\affiliation{\telaviv}

\date{\today}

\begin{abstract}
Lattice gauge theory is an important framework for studying gauge theories that arise in the Standard Model and condensed matter physics. 
Yet many systems (or regimes of those systems) are difficult to study using conventional techniques, such as action-based Monte Carlo sampling. 
In this paper, we demonstrate the use of gauged Gaussian projected entangled pair states as an ansatz for a lattice gauge theory involving dynamical physical matter. 
We study a $\mathbb{Z}_2$ gauge theory on a two dimensional lattice with a single flavor of fermionic matter on each lattice site. 
For small systems, our results show agreement with results computed by exactly diagonalizing the Hamiltonian, and demonstrate that the approach is computationally feasible for larger system sizes where exact results are unavailable. 
This is a further step on the road to studying higher dimensions and other gauge groups with manageable computational costs while avoiding the sign problem.
\end{abstract}
\maketitle

\section{Introduction}
\label{sec:introduction}

Gauge theories are central to the Standard Model of particle physics~\cite{peskin_introduction_1995}, as well as condensed matter physics~\cite{fradkin_field_2013}. 
Yet such theories are notoriously hard to study. Discretizing space (or spacetime) onto a lattice --- giving rise to a lattice gauge theory (LGT) --- provides a standard framework for studying gauge theories, particularly using Monte Carlo methods~\cite{wilson_confinement_1974, kogut_hamiltonian_1975, creutz_monte_1980}. 
However, doing so presents its own difficulties, prominent among them the sign problem, which arises when the probability intended for use in Monte Carlo does not form a valid probability distribution, i.e. takes on negative or complex values~\cite{troyer_computational_2005}.

We study a $\mathbb{Z}_2$ lattice gauge theory with gauge fields on the links and fermionic matter on the lattice sites.
Similar systems, including the pure $\mathbb{Z}_2$ gauge theory originally proposed by Wegner~\cite{wegner1971duality} and models with Ising matter fields~\cite{Fradkin_1979, tupitsyn2010}, as well as fermionic matter fields, have been studied widely in several contexts. 
These include $\mathbb{Z}_2$ spin liquids~\cite{savary2016quantum, Sachdev_2019}, unconventional quantum phases of matter~\cite{Senthil_2000,Prosko_2017, Gazit2017, gazit-confinement-2018, Cuadra_2020, Konig2020, Pozo_2021, borla_quantum_2022, brenig_spinless_2022}, and quantum simulation~\cite{Zohar_2017,barbiero2019,Homeier_2021, Lumia_2022, Homeier2023, Reinis2023,cochran2024visualizingdynamicschargesstrings}.
Here, we consider a $\mathbb{Z}_2$ theory in two spatial dimensions with a single flavor of fermionic matter.
This theory is a precursor to theories that suffer from a sign-problem and cannot be tackled with action-based Monte Carlo techniques.

Due to the above-mentioned computational problems, several methods have been proposed for the study of similar models in the last decade. 
One approach to the study of such systems is quantum simulation, with much recent progress and improvement as quantum hardware improves. 
Another is the use of tensor networks, and it is into this class which our work falls. Tensor networks are a class of ansatz states which are constructed by introducing extra ``virtual'' degrees of freedom, which are then traced out to produce a physical state~\cite{orus_practical_2014,cirac_matrix_2021}. They provide an important way of representing states based on entanglement properties, especially for states which obey an entanglement area law~\cite{hastings_area_2007,eisert_area_2010}.
Algorithms based on tensor networks are an important tool, since their scaling is often polynomial in the size of the system under consideration~\cite{white_density_1992, fannes_finitely_1992, schollwock_density-matrix_2011, cirac_matrix_2021}. 
Furthermore, tensor networks have been shown to capture the low energy states of many systems~\cite{white_density_1992, fannes_finitely_1992,hastings_area_2007}, and they are useful for studying thermal states~\cite{verstraete_matrix_2004}, as well as time evolution~\cite{daley_time-dependent_2004, zwolak_mixed-state_2004}.

The properties of tensor network states in one dimension, known as matrix product states (MPS), are relatively well-understood~\cite{cirac_matrix_2021}. In higher dimensions however, such as when working with projected entangled pair states (PEPS), which are the higher dimensional generalization of MPS, difficulties arise with both the theoretical and numerical aspects~\cite{schuch_computational_2007}. These issues largely have to do with the computational scaling of contracting the tensor networks and computing observables~\cite{cirac_matrix_2021,jordan_classical_2008,cirac_renormalization_2009}. 

Lattice gauge theories with tensor networks have been studied using several approaches. In a single space dimension, tensor network methods have been shown to overcome the sign problem and simulate time evolution well; details may be found in the review papers \cite{banuls_review_2020,banuls_simulating_2020} and references therein. Some numerical work has also been done in higher dimensions, such as the pure-gauge works \cite{tagliacozzo_entanglement_2011,tagliacozzo_tensor_2014}, as well as works using infinite PEPS (iPEPS) \cite{zapp_tensor_2017,robaina_simulating_2021} and tree tensor networks (TTN) \cite{felser_two-dimensional_2020,magnifico_lattice_2021,montangero_loop-free_2022,cataldi_21d_2023,magnifico2024roadmap}. 

While these methods are very powerful numerically, they are not especially tailored to physical gauge symmetries. PEPS, with their ability to describe global symmetries using a virtual local symmetry \cite{cirac_matrix_2021} suggest another path for studying lattice gauge theories in higher dimensions using tensor network states, building upon the concept of gauging, introduced in~\cite{haegeman_gauging_2015} and~\cite{zohar_building_2016}, through which globally invariant PEPS are made physically gauge invariant locally. We focus on the gauging procedure introduced in the latter reference, which was proven to be general enough for lattice gauge theory PEPS \cite{Kull2017ClassificationOM,blanik2024internalstructure}, and in particular for constructing gauged Gaussian projected entangled pair states. Note that another symmetry-tailored method is that of tensor renormalization group (see, e.g., the review \cite{meurice_tensor_2022}), but it focuses on path integral methods while we aim here at the Hamiltonian formalism with tensor network states.

Gauged Gaussian projected entangled pair states (GGPEPS) were introduced as a class of ansatz states designed for lattice gauge theories~\cite{zohar_fermionic_2015, zohar_projected_2016}. 
They build on the work of~\cite{Wahl_2014} which defined (ungauged) Gaussian PEPS.
GGPEPS provide a convenient way to ensure that the local constraints following from gauge invariance are met. As PEPS, they satisfy an entanglement area law --- the entanglement between two subsystems is proportional to the area (or lower/higher dimensional analogues) of the boundary between them. This is valuable, as the ground states of many LGTs of interest are known (in 1D) or conjectured (in higher dimensions) to obey such a law~\cite{eisert_area_2010}. GGPEPS also allow for the efficient computation of observables, including standard observables such as Wilson loops or mesonic operators as well as gradients of operators, which is important for calculating the gradients of the Hamiltonian for a ground state search via a minimization procedure~\cite{zohar_combining_2018}.

In this paper, we present results using gauged Gaussian PEPS for a theory with dynamical fermionic matter. 
In reference~\cite{kelman_2024} the theory surrounding these states
was developed in a general, abstract, and rigorous fashion, and in references~\cite{emonts_variational_2020} and~\cite{emonts_finding_2023}, these states were used to investigate the ground states of pure-gauge theories. This paper improves upon those by demonstrating numerical results with dynamic fermionic matter.
While the particular theory we consider does not suffer from the sign problem, it serves as a further demonstration that this approach offers a viable method for studying such theories as we work towards models that do suffer from the sign problem.

In the next section we describe the physical system we use for illustration --- a $\mathbb{Z}_2$ lattice gauge theory with dynamical fermionic matter. In the following section, we present our ansatz state and algorithm for finding the ground state. In the \nameref{sec:results} section
we present our results, including arguments that the results capture the physics of the desired state. We conclude with some reflections and plans for future work in the \nameref{sec:conclusions} section.

\section{Methods}
\label{sec:methods}

\subsection{Physical System}
\label{sec:system}

\begin{figure}[h]
	\includegraphics[width=0.7\linewidth]{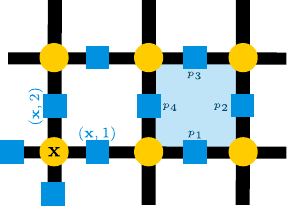}
	\caption{A diagram of the 2D lattice. Matter is shown on lattice sites in yellow and gauge fields on links in blue. The blue box shows the labeling convention for a plaquette.}
    \label{fig:lattice}
\end{figure}

A lattice gauge theory, such as the $\mathbb{Z}_N$ one we study here, has matter (fermionic, in our case) on the lattice sites and $\mathbb{Z}_N$ gauge fields on the links. 
We consider a two dimensional lattice of size $L_x$ by $L_y$ with periodic boundary conditions, and describe each lattice site by $\site{x} = (x, y) \in \mathbb{Z}^2$ and each link by $\ell = (\site{x}, k)$, where $k \in \{ 1,2\}$ denotes the direction of the link attached to the site $\site{x}$. This is illustrated in figure~\ref{fig:lattice}.
We label unit vectors in each direction by $\uvec{e}_1$ and $\uvec{e}_2$. 
It is occasionally convenient to refer to all the links around a given site; we therefore allow $k \in \{ 1,2,3,4\}$ and identify link $\ell = (\site{x}, 3)$ with the link $(\site{x} - \uvec{e}_1, 1)$ for horizontal links, and $\ell = (\site{x}, 4)$ with $(\site{x} - \uvec{e}_2, 2)$ for vertical links. 
It will be necessary to treat sites on the even and odd sublattices differently; we therefore define $(-1)^\site{x} = (-1)^{x + y}$ where $x, y$ are the coordinates of the site $\site{x}$, so that $(-1)^\site{x} = 1$ on the even sublattice and $(-1)^\site{x} = -1$ on the odd sublattice. 

The gauge fields are represented by group elements $g \in \mathbb{Z}_N$, which we use to construct an $N$-dimensional Hilbert space on each link. On each such Hilbert space we choose two unitary operators, $P$ and $Q$, which satisfy
\begin{equation} \begin{aligned}
\label{eq:Z2-algebra}
    P^N = Q^N = \id \\
    P^\dagger P = Q^\dagger Q = \id \\
    P Q P^\dagger = e^{i\delta}Q,
\end{aligned} \end{equation}
where $\delta = 2\pi/N$.  We denote the eigenstates of $P$ by $\{ \ket{p} \}$, which have eigenvalue $e^{ip\delta}$, and the eigenstates of $Q$ by $\{ \ket{q} \}$ which have eigenvalue $e^{iq\delta}$. $Q$ is then a raising operator on $\{ \ket{p} \}$ while $P$ is a lowering operator on $\ket{q}$, where the basis states are periodic, wrapping back to zero at $p=N$ or $q=N$~\cite{horn_hamiltonian_1979}.

We will focus here on the case of a $\mathbb{Z}_2$ theory to demonstrate the method.
In this case, these operators can be represented by Pauli matrices, for example $P = \sigma_z = P^\dagger$ and $Q = \sigma_x = Q^\dagger$.

We consider a single flavor of fermionic matter, without spin or color; to reduce the doubling problem \cite{nielsen_no-go_1981} we stagger \cite{susskind_lattice_1977} --- placing the two-components of the matter/anti-matter spinor on even/odd lattice sites respectively. A fermion on site $\site{x}$ is created by the operator $\psi^\dagger(\site{x})$, which obeys the anti-commutation relation for fermions, $\{ \psi^\dagger(\site{x}), \psi(\site{y}) \} = \delta(\site{x} , \site{y})$, where $\delta$ is the Kronecker delta.

The system is described by a Hamiltonian composed of terms which act on the gauge fields alone, the matter alone, and on both together~\cite{kogut_hamiltonian_1975}. For a $\mathbb{Z}_N$ gauge theory, the terms which depend solely on the gauge fields are given by \cite{horn_hamiltonian_1979} as 
\begin{equation} \begin{aligned}
\label{eq:pg-energy}
    H_\text{E} &= \sum_\ell (2 - (P_\ell + P_\ell^\dagger)) \\
    H_\text{B} &= \sum_{p} 2 (1 - Q_1 Q_2 Q_3^\dagger Q_4^\dagger), \\
\end{aligned} \end{equation}
and for $\mathbb{Z}_2$ specifically, this becomes
\begin{equation} \begin{aligned}
\label{eq:pg-energy-z2}
    H_\text{E} &= \sum_\ell 2 (1 - P_\ell) \\
    H_\text{B} &= \sum_{p} 2 (1 - Q_1 Q_2 Q_3 Q_4), \\
\end{aligned} \end{equation}
since $P$ and $Q$ are Hermitian in the $\mathbb{Z}_2$ theory. The offsets ensure that these terms are positive; they do not affect the physics. The subscripts $\text{E}$ and $\text{B}$ stand for ``electric'' and ``magnetic'', where these terms are borrowed from the $U(1)$ theory of electromagnetism. In the $U(1)$ theory, the analogous terms have that meaning (and in the large $N$ limit, compact QED, a lattice $U(1)$ model \cite{kogut_introduction_1979}, is obtained from the $\mathbb{Z}_N$ Hamiltonian \cite{horn_hamiltonian_1979}). As in that case, the electric term acts locally on the gauge fields, while the magnetic term is related to loops. The first term is a sum over all links, while the second is over all plaquettes (we have labelled the $Q$ operators by their position in the plaquette rather than with links labelled by site and direction, as shown in figure~\ref{fig:lattice}).

Next, we add a term acting solely on the matter degrees of freedom
\begin{equation} \begin{aligned}
\label{eq:mass-energy}
    H_\text{M}
        &= \frac{L_x L_y}{2} + \sum_{\site{x}} (-1)^{\site{x}} \psi^\dagger(\site{x}) \psi(\site{x}) \\
        &= \sum_{\site{x}} \Big( \frac{1}{2} + (-1)^{\site{x}} \psi^\dagger(\site{x}) \psi(\site{x})\Big), \\
\end{aligned} \end{equation}
where again the offset bounds the term below by zero. These are staggered fermions \cite{susskind_lattice_1977}, chosen for simplicity, where matter is placed on the even sites and anti-matter on the odd sites.

Finally we introduce an interaction term, which acts on each pair of neighboring sites together with the gauge field between them,
\begin{equation} \begin{aligned}
\label{eq:interaction-energy}
    H_\text{I}
        = 
        &\sum_{\site{x} }\Big( \eta^2 \psi^\dagger(\site{x}) U(\site{x},1) \psi(\site{x} + \uvec{e}_1) + h.c. \Big) \\
        - &\sum_{\site{x} }(-1)^{\site{x}} \Big( \psi^\dagger(\site{x}) U(\site{x},2) \psi(\site{x} + \uvec{e}_2) + h.c. \Big). \\
\end{aligned} \end{equation}
The phases $\eta = e^{i\pi/4}$ and $(-1)^\site{x}$ relate to rotations, as discussed below.
The gauging operator $U$ which acts on the gauge fields is defined for a $\mathbb{Z}_N$ gauge theory as
\begin{equation} \begin{aligned}
\label{eq:physical-gauging-operator}
    U(\site{x}, k) &= \sum_{0 \leq q < N} e^{i \delta q} \ket{q}\bra{q},
\end{aligned} \end{equation}
which gives that $U(\site{x}, k) = Q(\site{x}, k) = \sigma_x(\site{x}, k)$ for $\mathbb{Z}_2$. 
For a general group other than $\mathbb{Z}_N$, the system is described by a Hamiltonian given in~\cite{kogut_hamiltonian_1975,kogut_lattice_1983}.

Unlike the other terms, the interaction energy is not bounded below by zero. Furthermore, it is not trivial to find it's minimal value, which in general is not proportional to the lattice size. For a $2 \times 2$ lattice, the lowest possible value of this term can be calculated analytically as $-4\sqrt{2}$. For larger systems, the minimal value can be found using band structure considerations, as detailed in Supplementary Information~\ref{sec:free-fermions}:~\nameref{sec:free-fermions}. 

The full Hamiltonian is thus 
\begin{equation} \begin{aligned}
\label{eq:hamiltonian}
    H &= g_\text{E} H_\text{E} + g_\text{B} H_\text{B} + g_\text{I} H_\text{I} + g_\text{M} H_\text{M},
\end{aligned} \end{equation}
where $g_K$ are the couplings of the individual terms. 
Since there is only one flavor of matter, we do not add a chemical potential term.
We denote the electric coupling $g_\text{E}$ by $\lambda$, and set the magnetic coupling to $\frac{1}{\lambda}$, in order to match the relationship that arises through the discretization procedure of a continuous quantum field theory (QFT) onto a square lattice. Without consideration of the origins of the Hamiltonian in QFT, the couplings could be taken to vary independently.

The Hamiltonian and the states in which we are interested are invariant under several symmetries. 
The most central is local gauge invariance under the operators 
\begin{equation}
\label{eq:gauge-operator}
    V(\site{x}) = (-1)^\site{x} P_{\site{x}, 1} P_{\site{x}, 2}
        P_{\site{x}, 3} P_{\site{x}, 4}
        e^{i \pi \psi^\dagger(\site{x}) \psi (\site{x})},
\end{equation}
i.e. $[H, V(\site{x})] = 0$ for all lattice sites $\site{x}$. Note that we have here used the fact that $P = P^\dagger$ for $\mathbb{Z}_2$. This symmetry divides the Hilbert space into sectors which can be labeled by the eigenvalues of the operators $V(\site{x})$. We study the sector satisfying 
\begin{equation}
\label{eq:no-static-charges}
    V(\site{x}) \ket{\psi} = \ket{\psi} \quad \forall \ \site{x},
\end{equation}
i.e. the sector with no static charges anywhere on the lattice.

Since there are no static charges, the system is invariant under translations by two sites --- it is not invariant under translations of one site due to the staggering, which leads to the factor of $(-1)^\site{x}$ in equation~\eqref{eq:gauge-operator}.

The Hamiltonian is also invariant under rotations; since we work on a square lattice, the allowed rotations are those which rotate the plane by multiples of $\pi/2$. Since fermions pick up a minus sign upon a full rotation, we must account for this in the prescription for the rotation of fermionic operators,
\begin{equation}
\label{eq:fermion-rotation}
    \psi^\dagger(\site{x}) \to 
        \begin{cases}
            \eta \psi^\dagger(\Lambda \site{x}) \quad \site{x} \ \text{even} \\
            \bar{\eta} \psi^\dagger(\Lambda \site{x})  \quad \site{x} \ \text{odd}
        \end{cases} \\
\end{equation}
where $\Lambda \site{x} = \Lambda(x,y) = (-y,x)$ implements the rotation, and $\eta = e^{i \pi /4}$. 
The operators on the links do not pick up a phase, and transform as 
\begin{equation} \begin{aligned}
\label{eq:link-rotation}
    Q(\site{x}, k) &\to 
        \begin{cases}
            Q(\Lambda \site{x}, 2) & k = 1 \\
            Q( \Lambda \site{x} - \uvec{e}_1, 1) & k = 2.
        \end{cases}
\end{aligned} \end{equation}
Only with rotations defined in this way is the Hamiltonian invariant under the allowed rotations.

Finally, the Hamiltonian is invariant under a global $U(1)$ symmetry, $\psi^\dagger(\site{x}) \to e^{i\theta} \psi^\dagger(\site{x})$ for all sites. This corresponds to a total number conservation, i.e. the symmetry is generated by the conserved quantity
\begin{equation}
    N = \sum_\site{x} \psi^\dagger(\site{x}) \psi(\site{x}).
\end{equation}
Since $[H, N] = 0$, this operator also divides the Hilbert space into sectors labelled by its eigenvalues; we focus on the ``half-filling'' case, in which there are equal numbers of matter and anti-matter particles. This sector is that of the Dirac sea, in which all even sites are empty while odd sites are full, and is given by
\begin{equation}
    N \ket{\psi} = \frac{L_x L_y}{2} \ket{\psi}.
\end{equation}

We are interested in finding the ground state energy of the Hamiltonian given in equation~\eqref{eq:hamiltonian} and describing its physics. To do so, we build an ansatz state and preform a variational Monte Carlo minimization over its variational parameters, as we describe in the next section.
Once a particular state of interest --- in our case, the ground state --- has been found, one may also be interested in observables other than the energy. Particular observables of interest include Wilson and Polyakov loops as well as mesonic operators. The closed loop operators take the form of products of link operators $U$ along the path determined by the loop; mesonic operators are similar, but with matter operators $\psi^\dagger$ and $\psi$ at the ends of the string. In both cases, the loop or string is a directed path, and one must consider the conjugate loop operators on those links which are traversed in the negative directions. In the case of the $\mathbb{Z}_2$ theory studied here, this slight complication can be ignored, since all loop operators are Hermitian.
In the~\nameref{sec:results} section we show some examples of Wilson loops calculations on the ground state.

\subsection{Ansatz \& Algorithm}

\subsubsection{Particle Hole Transformation}
\label{sec:particle-hole}

Before constructing the ansatz state, it is useful to perform a transformation that will simplify the construction, and make the resulting system and ansatz state translationally invariant under translations by a single site.
The transformation swaps the creation and annihilation operators on the odd sublattice, which allows us to treat the anti-matter on the odd sites in exactly the same way as the matter on the even sites. We therefore call it a ``particle-hole transformation.''
It is
\begin{equation} \begin{aligned}
    \psi^\dagger(\site{x})
        \to 
            \begin{cases}
            \tilde{\psi}^\dagger(\site{x}) & \site{x} \ \text{even} \\
            \tilde{\psi}(\site{x})    & \site{x} \ \text{odd}
            \end{cases}
    \\
    \psi(\site{x})
        \to 
            \begin{cases}
            \tilde{\psi}(\site{x})    & \site{x} \ \text{even} \\
            \tilde{\psi}^\dagger(\site{x})    & \site{x} \ \text{odd}.
            \end{cases}
\end{aligned} \end{equation}
Note that this mathematical transformation does not change the underlying physics.

After the transformation, the system is invariant under translations by a single site. The symmetry under rotations remains, but with the modification of equations~\eqref{eq:fermion-rotation} and~\eqref{eq:link-rotation} to
\begin{equation} \begin{aligned}
    \tilde{\psi}^\dagger(\site{x}) &\to \eta \tilde{\psi}^\dagger(\Lambda \site{x}), \\
    Q(\site{x}, k) &\to \begin{cases}
                    Q(\Lambda \site{x}, 2) & k = 1 \\
                    - Q( \Lambda \site{x} - \uvec{e}_1, 1) & k = 2.
                \end{cases}
\end{aligned} \end{equation}
Since $U(\site{x}, k) = Q(\site{x}, k)$, this also gives the transformation rules for $U(\site{x}, k)$.
The global $U(1)$ symmetry of total particle number conservation is also transformed, into
\begin{equation} \begin{aligned}
\label{eq:u1-after-ph}
    \tilde{\psi}^\dagger(\site{x}) 
        &\rightarrow_{U(1)}
            \begin{cases}
            e^{i\theta} \tilde{\psi}^\dagger(\site{x})  & \site{x} \ \text{even} \\
            e^{- i\theta} \tilde{\psi}^\dagger(\site{x})   & \site{x} \ \text{odd},
            \end{cases}
\end{aligned} \end{equation}
while gauge invariance is now defined as invariance under 
\begin{equation}
\label{eq:ph-gauge-operator}
    V(\site{x}) = P_{\site{x}, 1} P_{\site{x}, 2}
        P^\dagger_{\site{x}, 3} P^\dagger_{\site{x}, 4}
        e^{i \pi \tilde{\psi}^\dagger(\site{x}) \tilde{\psi} (\site{x})}.
\end{equation}

Finally, we must rewrite the Hamiltonian in terms of the transformed fermionic operators.
The operators which act only on the gauge fields remain unchanged, but the mass and interaction terms become
\begin{equation} \begin{aligned}
    H_\text{M}
        &= \sum_{\site{x}} \tilde{\psi}^\dagger(\site{x}) \tilde{\psi}(\site{x}), \\
    H_{\text{I}}
        &= \sum_{\site{x}} 
             \Big[ \eta^2 \tilde{\psi}^\dagger(\site{x}) Q(\site{x},1) \tilde{\psi}^\dagger(\site{x} + \uvec{e}_1) 
            \\ &\qquad \qquad \qquad
            + \bar{\eta}^2 \tilde{\psi}(\site{x} + \uvec{e}_1) Q(\site{x},1) \tilde{\psi}(\site{x}) \Big] \\
            &\qquad - \Big[ \tilde{\psi}^\dagger(\site{x}) Q(\site{x},2) \tilde{\psi}^\dagger(\site{x} + \uvec{e}_2) 
            \\ &\qquad \qquad \qquad
            + \tilde{\psi}(\site{x} + \uvec{e}_2) Q(\site{x},2) \tilde{\psi}(\site{x}) \Big]. \\
\end{aligned} \end{equation}
In the remainder of this paper, we drop the $\tilde{\cdot}$ on $\tilde{\psi}$.

\subsubsection{Ansatz}
\label{sec:ansatz}

A general state in the Hilbert space of the full system --- matter and gauge fields together --- is a ray that can be written
\begin{equation} \begin{aligned}
    \ket{\psi}
        &= \sum_{F, \mathcal{G}} \varphi(F, \mathcal{G}) \ket{F} \ket{\mathcal{G}}, \\
\end{aligned} \end{equation}
i.e. as a sum over basis states of the fermionic matter $\ket{F}$ and gauge fields $\ket{\mathcal{G}}$, where $\varphi(F, \mathcal{G})$ is a complex amplitude.
However, in general, such a state will not be gauge invariant --- to ensure gauge invariance one must impose conditions on $\varphi(F, \mathcal{G})$ which rule out non-gauge-invariant states.
Defining the gauge field configuration $\ket{\mathcal{G}}$ as 
\begin{equation} \begin{aligned}
\label{eq:gauge-field-configuration}
    \ket{\mathcal{G}}
        &= \bigotimes_{\site{x}, k} \ket{g(\site{x}, k}, \\
\end{aligned} \end{equation}
one can always write a state in the form
\begin{equation} \begin{aligned}
\label{eq:ansatz-integral-form}
    \ket{\psi}
        &= \sum_\mathcal{G} \ \psi_I(\mathcal{G}) \ket{\psi_{II}(\mathcal{G})} \ket{\mathcal{G}}, \\
\end{aligned} \end{equation}
where $\ket{\psi_{II}(\mathcal{G})}$ is a state of the matter, which now depends on the gauge field configuration in a way that ensures gauge invariance, and $\psi_I(\mathcal{G})$ is a complex amplitude. The sum is over all possible gauge field configurations on all the links, which we express in the basis of eigenstates of $U(\site{x}, k) = Q(\site{x}, k)$, denoted the magnetic basis (though equation~\eqref{eq:ansatz-integral-form} remains valid in any basis).

We wish to construct an ansatz for both $\psi_I(\mathcal{G})$ and $\ket{\psi_{II}(\mathcal{G})}$ which satisfies the desiderata already mentioned: it should be computationally tractable to evaluate observables, gauge invariant, and result in a final state which obeys an entanglement area law. The GGPEPS ansatz meets all of these conditions.

\begin{figure}[h]
	\includegraphics[width=1.0\linewidth]{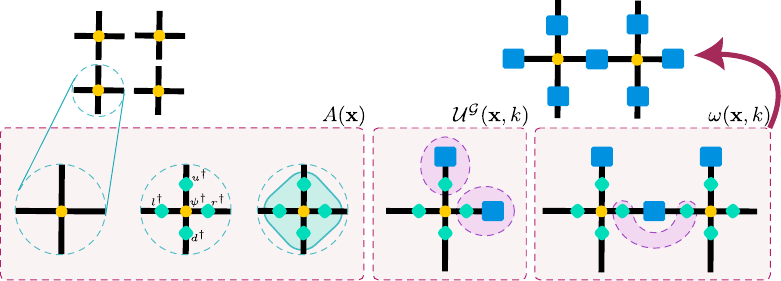}
	\caption{A graphical representation of the construction of a gauged gaussian projected entangled pair state (GGPEPS) for a fixed gauge field configuration. We start on the left with the vacuum of the physical modes on each site (shown in yellow), create 4 virtual modes on each link around each site (turquoise), and couple them to each other and to the physical modes. We then couple the appropriate virtual modes to the gauge fields (blue). Finally, we project the virtual modes from neighboring sites onto a maximally entangled state, and trace out the virtual modes.
    The bottom row illustrates each operation locally; it is in fact applied to the entire lattice at each stage.}
    \label{fig:ggpeps-construction}
\end{figure}

We construct $\psi_I(\mathcal{G})$ and $\ket{\psi_{II}(\mathcal{G})}$  as fermionic Gaussian PEPS \cite{kraus_fermionic_2010} in very similar ways (in the language of~\cite{kelman_2024}, they are simply different layers of the ansatz), but with the inclusion of physical matter for $\ket{\psi_{II}(\mathcal{G})}$. A graphical representation of the construction is shown in figure~\ref{fig:ggpeps-construction}.

As in many tensor-network approaches, we begin by introducing virtual modes --- for each site, in each direction, we add new degrees of freedom. We label these modes by the link direction with which they are associated $k \in \{1,2,3,4\}$ as well as an index $\mu$ to indicate the ``copy'', since we add multiple virtual modes per link per site. In our case, we add four modes per link per site (for reasons discussed in Supplementary Information~\ref{sec:ansatz-details}:~\nameref{sec:ansatz-details}), giving a total of 16 virtual modes per site created by $\alpha^\dagger_{k\mu}(\site{x})$. For convenience, we will often denote the modes with a single index that wraps $k$ and $\mu$, and will include the physical mode $\psi^\dagger(\site{x})$ among them.

The new virtual modes are coupled to each other and to the physical matter associated with the appropriate site through an operator of the Gaussian form
\begin{equation}
\label{eq:A-op}
    A(\site{x}) = \exp \Big( \mathcal{T}_{ij}(\site{x}) 
        \alpha^\dagger_i(\site{x}) \alpha^\dagger_j(\site{x}) \Big),
\end{equation}
where $i,j$ run over all the physical and virtual modes, which we denote by $\alpha$. Summation is implied over $i,j$. This operator acts on the product of the physical and virtual Fock vacua $\ket{\Omega_\text{p}}, \ket{\Omega_\text{v}}$.
Since there is one physical mode per site, and we have added $4(2d) = 16$ (where $d=2$ is the dimension of the lattice) virtual modes, $\mathcal{T}$ is a $17 \times 17$ matrix.
In order to ensure the final state obeys all the symmetries of the system discussed above, we must impose constraints on $\mathcal{T}$ --- the details of how to do so are in Supplementary Information~\ref{sec:ansatz-details}:~\nameref{sec:ansatz-details}. Of the four virtual modes per site per link, two will be used in the construction of $\psi_I(\mathcal{G})$ and two in the construction of $\ket{\psi_{II}(\mathcal{G})}$. These sets of modes will not couple to each other, and the physical mode will only couple to the virtual modes used in the construction of $\ket{\psi_{II}(\mathcal{G})}$, so $A(\site{x})$ can be decomposed as
\begin{equation}
\label{eq:A-decomposition}
    A(\site{x}) = A_I(\site{x}) A_{II}(\site{x}),
\end{equation}
where each $A_J(\site{x})$ has the form given in equation~\eqref{eq:A-op}, involving only some of the modes --- for $A_I(\site{x})$, two copies of virtual modes per link (for a total of 8); for $A_{II}(\site{x})$, two other copies of virtual modes per link plus the single physical mode on the site $\site{x}$ (for a total of 9).

Next we introduce a gauging operator, following the procedure of \cite{zohar_building_2016} which was proven to be general in~\cite{Kull2017ClassificationOM, blanik2024internalstructure}. This operator couples the gauge field on each link to the virtual modes just introduced. The coupling guarantees that the state will be gauge invariant. It is sufficient to only couple the gauge field on a link to the virtual modes associated with one of the neighboring sides; we couple the field on each link to the virtual modes associated with the site $\site{x}$ when the link is represented as $\ell = (\site{x}, k)$ with $k \in \{1,2\}$, i.e. the site to the left or below the given link.
For $\mathbb{Z}_2$, the gauging operator 
can be written as 
\begin{equation} \begin{aligned}
\label{eq:gauging-op}
    \mathcal{U}^\mathcal{G}(\site{x}, k)
        &= \id \otimes \mathds{P}_{\text{even}} 
        + Q(\site{x}, k) \otimes \mathds{P}_{\text{odd}} \\
\end{aligned} \end{equation}
where here $\alpha_i$ ranges over the virtual modes of link $(\site{x}, k)$ which arise from $\site{x}$ (but not those which arise from site $\site{x} \pm \uvec{e}_k$). In writing equation~\eqref{eq:gauging-op}, we have made use of equation~\eqref{eq:physical-gauging-operator}, which gives that $U(\site{x}, k) = Q(\site{x}, k)$.
$\mathds{P}_{\text{even}}$ is a projection operator onto the state with an even number of excited virtual modes; $\mathds{P}_{\text{odd}}$ is defined analogously.
This gauging operator can also be decomposed into operators which act only on the virtual modes used in the construction of $\psi_I(\mathcal{G})$ and $\ket{\psi_{II}(\mathcal{G})}$, so that 
\begin{equation}
\label{eq:gauging-decomposition}
    \mathcal{U}^\mathcal{G}(\site{x}, k) 
        = \mathcal{U}_I^\mathcal{G}(\site{x}, k) 
        \mathcal{U}_{II}^\mathcal{G}(\site{x}, k),
\end{equation}
where each of $\mathcal{U}_J^\mathcal{G}(\site{x}, k)$ are defined as in equation~\eqref{eq:gauging-op} over the appropriate virtual modes. Note that both $\mathcal{U}_J^\mathcal{G}(\site{x}, k)$ operate on the same --- physical --- gauge field.

The final step in the construction of the ansatz state is the projection of the virtual modes from neighboring sites (on the same link) onto a maximally entangled state, and tracing out the virtual modes to leave a state with only physical degrees of freedom (the physical matter and the gauge fields).
We define
\begin{equation} \begin{aligned}
\label{eq:projector-def}
    w(\site{x}, k) &= 
    \exp \Big( W^{(k)}_{ij}
    \alpha^{\dagger}_i(\site{x})
    \alpha^{\dagger}_j(\site{x} + \uvec{e}_k)
    \Big)
\end{aligned} \end{equation} 
with summation implied over $i,j$. $W^{(k)}_{ij}$ determines the coupling between the virtual modes. 
It is convenient to define notation for the virtual modes such that $r^\dagger_\mu, u^\dagger_\mu, l^\dagger_\mu, d^\dagger_\mu$ as those which are right, up, left, down of the site, as shown in figure~\ref{fig:ggpeps-construction}. Thus
\begin{equation} \begin{aligned}
    r^\dagger_\mu(\site{x}) &= \alpha^\dagger_{1\mu}(\site{x}) \qquad
    u^\dagger_\mu(\site{x}) &= \alpha^\dagger_{2\mu}(\site{x}) \\
    l^\dagger_\mu(\site{x}) &= \alpha^\dagger_{3\mu}(\site{x}) \qquad
    d^\dagger_\mu(\site{x}) &= \alpha^\dagger_{4\mu}(\site{x}). \\
\end{aligned} \end{equation}
For the particular ansatz used in the results below, we choose $W^{(k)}$ to give
\begin{equation} \begin{aligned}
\label{eq:projector-operators-layer1}
    w_I(\site{x}, 1)
        &= \exp \big( r_1(\site{x}) l_2(\site{x} + \uvec{e}_1) 
            + r_2(\site{x}) l_1(\site{x} + \uvec{e}_1) \big) \\
    w_I(\site{x}, 2)
        &= \exp \big( \eta^2 u_1(\site{x}) d_2(\site{x} + \uvec{e}_2) 
            + \eta^2 u_2(\site{x}) d_1(\site{x} + \uvec{e}_2) \big) \\
\end{aligned} \end{equation}
for $\psi_I(\mathcal{G})$, and 
\begin{equation} \begin{aligned}
\label{eq:projector-operators-layer2}
    w_{II}(\site{x}, 1)
        &= \exp \big( r_3(\site{x}) l_3(\site{x} + \uvec{e}_1) 
            + r_4(\site{x}) l_4(\site{x} + \uvec{e}_1) \big) \\
    w_{II}(\site{x}, 2)
        &= \exp \big( \eta^2 u_3(\site{x}) d_3(\site{x} + \uvec{e}_2) 
            + \eta^2 u_4(\site{x}) d_4(\site{x} + \uvec{e}_2) \big) \\
\end{aligned} \end{equation}
for $\ket{\psi_{II}(\mathcal{G})}$, which together give
\begin{equation}
\label{eq:projector-decomposition}
    w(\site{x}, k) = w_I(\site{x}, k) w_{II}(\site{x}, k).
\end{equation}
The symmetries under which the projecting operators are invariant are discussed in Supplementary Information~\ref{sec:ansatz-details}:~\nameref{sec:ansatz-details}; see also~\cite{kelman_2024}. 

With all of these ingredients we define
\begin{equation}
    \psi_I(\mathcal{G}) = 
    \bra{\Omega_I} \prod_{\site{x},k} w_I(\site{x},k) \prod_{\site{x},k} \mathcal{U}_I^\mathcal{G}(\site{x},k) \prod_{\site{x}} A_I(\site{x}) \ket{\Omega_I}
\end{equation}
and 
\begin{equation} \begin{aligned}
    \ket{\psi_{II}(\mathcal{G})} &= 
    \bra{\Omega_{II}} \prod_{\site{x},k} w_{II}(\site{x},k) \prod_{\site{x},k} \mathcal{U}_{II}^\mathcal{G}(\site{x},k) 
    \\ & \qquad \qquad
    \prod_{\site{x}} A_{II}(\site{x}) \ket{\Omega_{II}} \ket{\Omega_\text{p}} 
\end{aligned} \end{equation}
where $\ket{\Omega_J}$ is the Fock vacuum for the virtual modes used in each construction; $\ket{\Omega_\text{v}} = \ket{\Omega_I} \ket{\Omega_{II}}$. Similarly, $\ket{\Omega_\text{p}}$ is the Fock vacuum of the physical matter.
The first two products are over all links while the last is over all sites.
If we ignore the portion of $U_{J}^\mathcal{G}$ which consists of an operator on the gauge fields, $\psi_I(\mathcal{G})$ is a complex number while $\ket{\psi_{II}(\mathcal{G})}$ is a state of the matter fields.

Substituting into equation~\eqref{eq:ansatz-integral-form}, the full ansatz can be written
\begin{equation} \begin{aligned}
    \ket{\psi}
        &= \sum_\mathcal{G} \ \bra{\Omega_\text{v}} \prod_{\site{x},k} \omega(\site{x},k) \prod_{\site{x},k} \mathcal{U}^\mathcal{G}(\site{x},k) 
        \\ & \qquad \qquad
        \prod_{\site{x}} A(\site{x}) \ket{\Omega_\text{v}} \ket{\Omega_\text{p}} \ket{\mathcal{G}} \\
\end{aligned} \end{equation}
which satisfies all that we wished from a state: guaranteed gauge invariance, an entanglement area law, and efficient representation for calculation of observables. Note that the only variational parameters of the state are contained in $A$ by way of $\mathcal{T}$. 

We finish this section with a brief review of some properties of this state.
We first note that for any particular gauge field configuration $\mathcal{G}$, the state $\psi_I(\mathcal{G}) \ket{\psi_{II}(\mathcal{G})}$ is Gaussian. 
As a result, this fermionic state is fully characterized by its covariance matrix, which can be computed efficiently, and from which (together with the gauge field configuration) all observables can be calculated.
We therefore satisfy the first requirement --- computationally efficient calculation of observables. 

\begin{figure}[h]
	\includegraphics[width=0.7\linewidth]{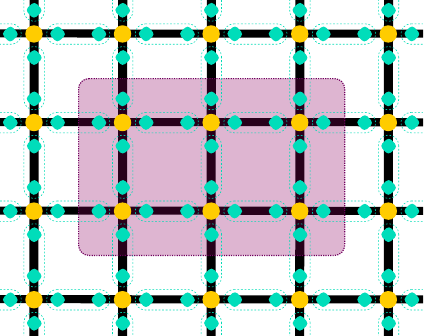}
	\caption{A graphical representation of the state which shows that an entanglement area law is automatically obeyed. The physical matter on the lattice sites is shown in yellow, the virtual modes are shown in turquoise, and the gauge fields are not shown. The entanglement between the region highlighted in purple and the rest of the lattice is proportional to the number of links through which the boundary cuts, since the only source of entanglement is the projection onto maximally entangled states of the virtual modes on the same link (enclosed by the turquoise dashed lines).}
    \label{fig:area-law}
\end{figure}

It is known that local, gapped Hamiltonians in one dimension obey an entanglement area law --- that is, the entanglement between subsystems grows proportionally to the size of the boundary between them, which in one dimension is simply a constant~\cite{hastings_area_2007, eisert_area_2010}.
The analogous statement is conjectured rather than proven in two dimensions, but our ansatz satisfies this property automatically --- the entanglement between subsystems can arise only due to the entanglement introduced by the projection operators $w(\site{x}, k)$, and along any curve which divides the lattice into two, the number of such operators is proportional to the length of the curve~\cite{kelman_2024}. This is illustrated in figure~\ref{fig:area-law}.

Finally, our state is gauge invariant --- it is an eigenstate of the operator defined in equation~\eqref{eq:gauge-operator}. In particular, it satisfies equation~\eqref{eq:no-static-charges} (or equivalently equation~\eqref{eq:ph-gauge-operator}, which defines gauge-invariance for the transformed fermionic operators), though the ansatz is general enough to handle other sectors as well~\cite{kelman_2024}.

\subsubsection{Algorithm}
We wish to study the system described in the ~\nameref{sec:system} subsection of the~\nameref{sec:methods} section by finding its ground state.
To do so, we employ a variational Monte Carlo procedure. The key insight is that the expectation value of any observable can be written as
\begin{equation} 
\label{eq:obs-expectation}
    \left\langle \mathcal{O} \right\rangle = \sum_\mathcal{G} F_{\mathcal{O}}\left(\mathcal{G}\right)p\left(\mathcal{G}\right),
\end{equation}
where the summation is over all possible gauge field configurations (in the results presented here, we do not make use of gauge fixing).  $F_{\mathcal{O}}\left(\mathcal{G}\right)$ can be expressed in terms of the covariance matrices of the state, and 
\begin{equation} \begin{aligned}
\label{eq:probability}
    p(\mathcal{G})
        &= \frac{ \abs{\psi_I(\mathcal{G})}^2 \braket{\psi_{II}(\mathcal{G})} }
        { \sum_{\mathcal{G}'} \abs{\psi_I(\mathcal{G}')}^2 \braket{\psi_{II}(\mathcal{G}')} }
\end{aligned} \end{equation}
defines a valid probability distribution that can be used for Monte Carlo.
The expressions for $F_{\mathcal{O}}\left(\mathcal{G}\right)$ for the electric and magnetic terms of the energy are given in~\cite{emonts_finding_2023}. The expressions for the interaction and mass terms are given in Supplementary Information~\ref{sec:observables}:~\nameref{sec:observables}.

To find the ground state, the relevant observables are the gradients of the energy --- in particular, the gradients with respect to the parameters that define a particular state, which are all contained in $\mathcal{T}(\site{x})$ in $A(\site{x})$.
We start by initializing the state with random parameters (subject to the constraints of Supplementary Information~\ref{sec:ansatz-details}:~\nameref{sec:ansatz-details}), evaluate the gradients with Monte Carlo, and then update the parameters using the BFGS optimization algorithm until convergence is reached.

It is also worth noting that in the Monte Carlo procedure, we must calculate observables for many gauge field configurations, but that this does not require calculating the covariance matrix from scratch. Instead local updates can be applied at intermediate steps in the calculation depending on which gauge fields were modified \cite{kelman_2024}. This further improves the computational efficiency of this approach.

\section{Results}
\label{sec:results}

\begin{figure}[h]
    \includegraphics[width=1.0\linewidth]{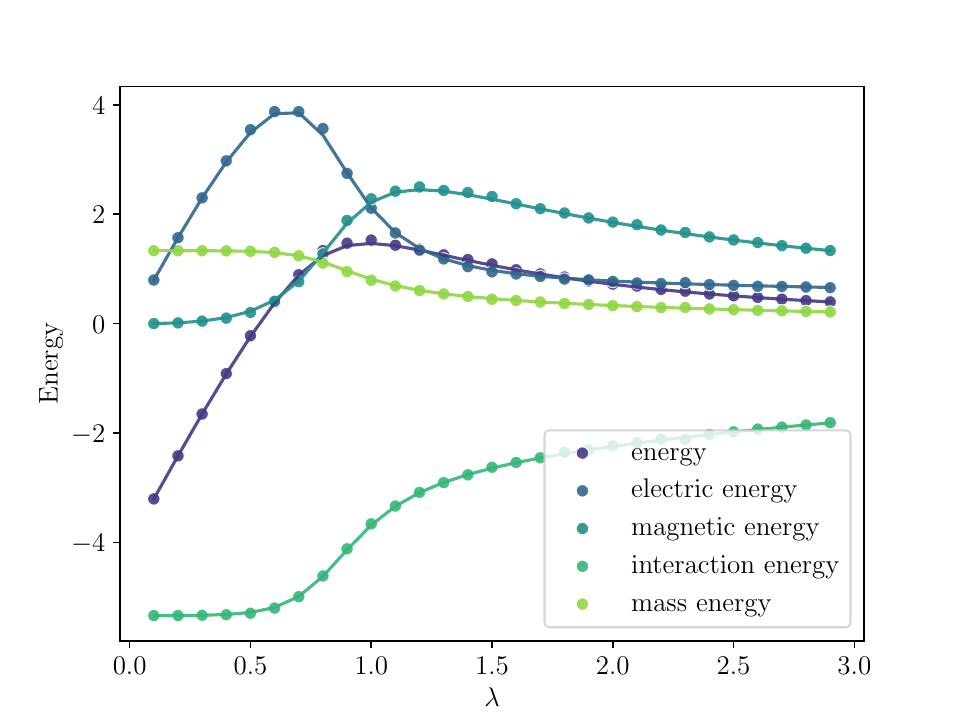}
	\caption{The various energy observables for the $2\times2$ ground states. The dots show results given by our ansatz, showing their close match to the exact diagonalization results (solid lines) for each term in the Hamiltonian. The interaction coupling $g_\text{I}$ was fixed at $g_\text{I} = 1.0$, and the mass coupling $g_\text{M}$ was fixed at $g_\text{M} = 1.0$. The horizontal axis is the electric coupling $g_\text{E} = \lambda$; the magnetic coupling is defined as $g_\text{B} = 1/\lambda$. The results shown here do not rely on Monte Carlo, but rather on exact contraction --- for a $2 \times 2$ system, it is possible to simply iterate over all $2^{2L^2} = 256$ gauge configurations of the entire lattice.
    No error bars are shown as no sampling error is present and we have not quantified numerical error due to the minimization procedure.}
    \label{fig:energy-observables}
\end{figure}

In the previous sections we described our ansatz state and minimization procedure for finding the ground state of the $\mathbb{Z}_2$ Kogut-Susskind Hamiltonian shown in equation~\eqref{eq:hamiltonian}. Some details regarding the configuration of Monte Carlo and the minimization are given in Supplementary Information~\ref{sec:computation-details}:~\nameref{sec:computation-details}.
As a first check of our method, we compare our results with results found by exactly diagonalizing the Hamiltonian. This could only be done (under practical time and memory constraints) for a $2 \times 2$ and $4\times4$ lattice. Note that due to the staggering of matter, the extent of the lattice in each dimension must be an even number.
Figure~\ref{fig:energy-observables} shows a scan of the electric (and magnetic) couplings, with the corresponding ground state energy as well as each term of the energy --- compared with exact diagonalization results.

The most important result from figure~\ref{fig:energy-observables} is that our ansatz not only finds the ground state energy, but also correctly describes the physics of the ground state, as seen from the close agreement of each term in the energy with the exact results.
Note that since the state is found using a variational procedure it provides an upper bound on the total energy of the true ground state, but there is no such guarantee on each term.

\begin{figure}[h]
	\includegraphics[width=1.0\linewidth]{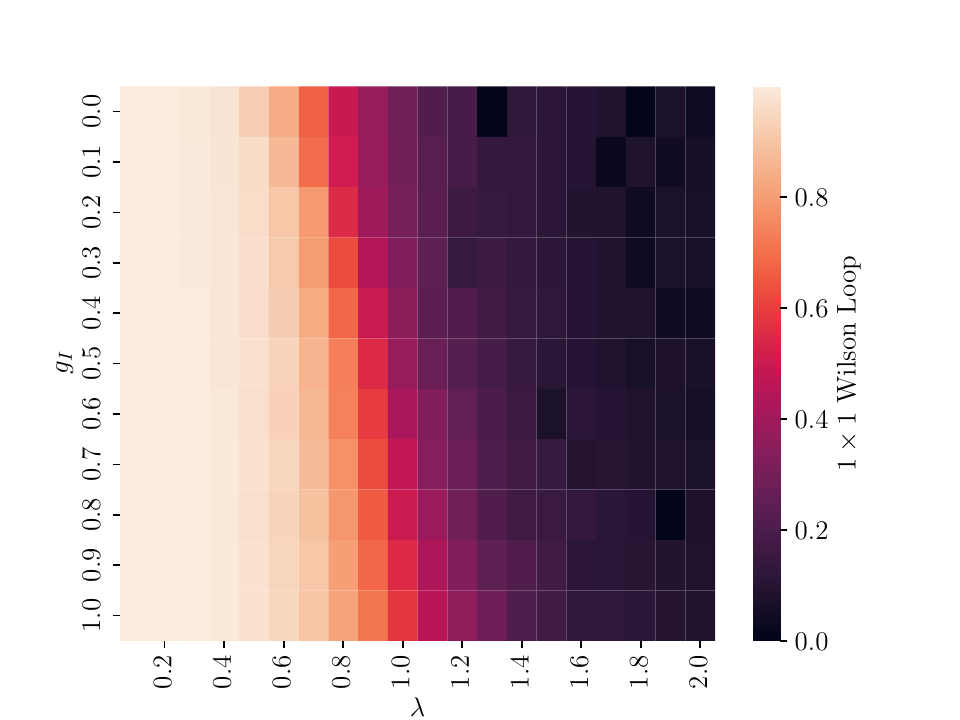}
	\caption{A parameter diagram for a $2\times2$ system with massless fermions ($g_\text{M} = 0$) coupled to gauge fields. The horizontal axis is the electric coupling $g_\text{E} = \lambda$ (the magnetic coupling is given by $1/\lambda$), and the vertical axis is the interaction coupling $g_\text{I}$. The colors indicate the value of $1\times1$ Wilson loops for the ground state of the Hamiltonian with the specified couplings.}
    \label{fig:wilson_L2}
\end{figure}

The couplings in the Hamiltonian (equation~\eqref{eq:hamiltonian}) define a three dimensional parameter space (since an overall energy scaling is irrelevant, we can fix one of the couplings).
A two dimensional slice of this parameter space is shown in figure~\ref{fig:wilson_L2}, with the results again calculated using exact contraction for a $2\times2$ system.
The figure shows the value of $1\times1$ Wilson loops. Though Wilson loops are not an order parameter when fermionic matter is included, the transition between high and low values perhaps signals a phase transition between confined and unconfined phases; we leave an investigation of phases for future work.

\begin{figure*}
	\includegraphics[width=1.0\linewidth]{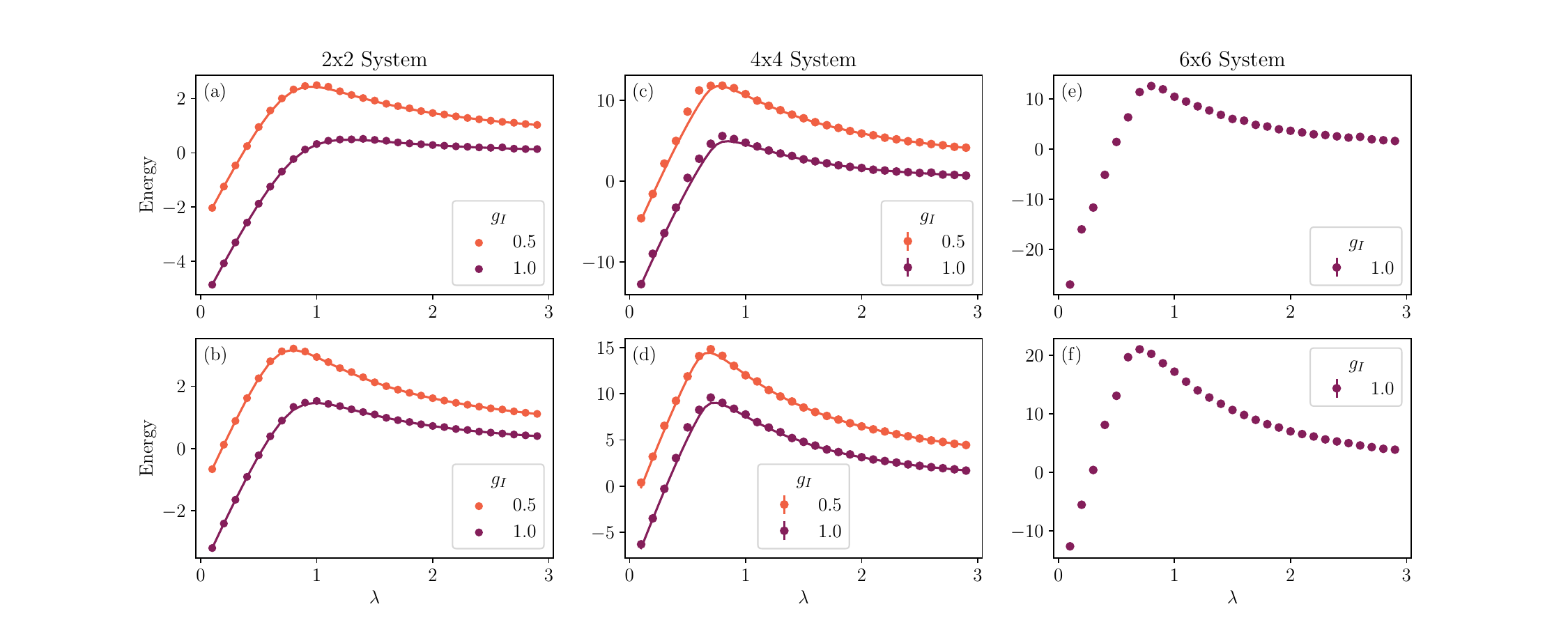}
	\caption{The ground state energies found for various lattice sizes and couplings. Moving from left to right, panels (a-b) show a $2\times2$ lattice (computed via exact contraction), panels (c-d) show a $4\times4$ lattice (computed with Monte Carlo), and panels (e-f) show a $6\times6$ lattice (computed with Monte Carlo). Where available, the results computed via exact diagonalization are shown as solid lines.
    The top row (panels a, c, e) is systems with $g_\text{M} = 0$ (massless fermions), while the bottom row (panels b, d, f) gives results for $g_\text{M} = 1$. The interaction coupling $g_\text{I}$ is shown in the legend, while $\lambda$ on the $x$-axis defines the electric and magnetic couplings ($g_\text{E} = \lambda$ and $g_\text{B} = 1/\lambda$).
    The error bars shown in the legend (which are generally too small to be seen on the graph) are the errors due to Monte Carlo sampling; error due to imperfect minimization is not shown.}
    \label{fig:energy-plots}
\end{figure*}

Figure~\ref{fig:energy-plots} shows the ground state energy for lattice sizes $2 \times 2$, $4\times4$, and $6\times6$ as a function of all of the couplings. The results for the $2 \times 2$ were calculated using exact contraction, while the $4\times4$ and $6\times6$ results used Monte Carlo. The top row shows massless fermions ($g_\text{M} = 0$) while the bottom row shows massive fermions ($g_\text{M} = 1.0$).
The results in the upper left pane ($2\times2$ lattice with massless fermions) correspond to horizontal slices of figure~\ref{fig:wilson_L2}; the results in the lower left pane ($2\times2$ lattice with massive fermions) with $g_\text{I} = 1.0$ are the same as the total energy shown in figure~\ref{fig:energy-observables}. The $4\times 4$ exact-diagonalization data (which required $>100$ GB of memory for each point) makes it clear that the ansatz has some difficulty with the optimization at the peak energy, though with greater computation time, the convergence could likely be improved.

\begin{figure}[h]
	\includegraphics[width=1.0\linewidth]{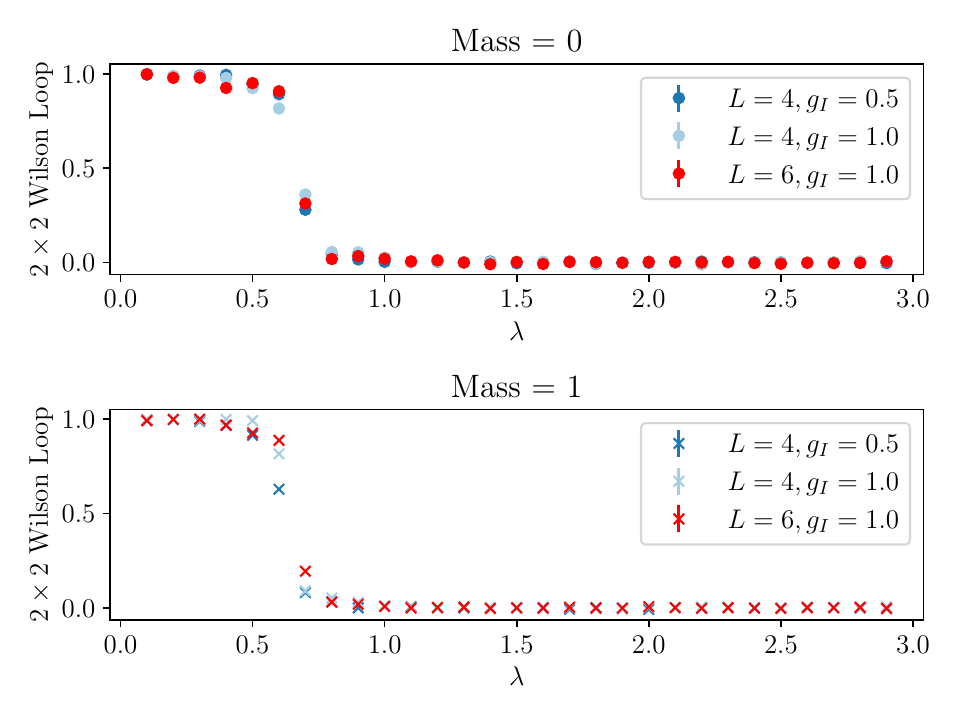}
	\caption{The expectation values of $2\times2$ Wilson loops for $4\times4$ and $6\times6$ systems. 
    Though Wilson loops are not an order parameter for theories including matter, the sharp transition is noteworthy. 
    The interaction coupling $g_\text{I}$ is shown in the legend, while $\lambda$ on the $x$-axis defines the electric and magnetic couplings ($g_\text{E} = \lambda$ and $g_\text{B} = 1/\lambda$).
    The error bars shown in the legend (which are generally too small to be seen on the graph) are the errors due to Monte Carlo sampling.}
    \label{fig:wilson-larger-systems}
\end{figure}

Finally, we demonstrate that our method can be extended to larger lattice sizes, which are beyond the reach of exact diagonalization, and that we capture the physics of Wilson loops for such systems. Figure~\ref{fig:wilson-larger-systems} shows Wilson loops for $4\times4$ and $6\times6$ systems. Here too we see a sharp change in the values of Wilson loops across the transition in ground state energies shown in figure~\ref{fig:energy-plots}.

\section{Discussion}
\label{sec:conclusions}

This paper demonstrated the use of GGPEPS for studying a lattice gauge theory with dynamical fermions. Previous work has shown that GGPEPS can capture the low energy physics of pure-gauge $\mathbb{Z}_2$ and $\mathbb{Z}_3$ theories.
The key advance demonstrated here is the inclusion of fermionic matter --- in addition to gauge fields --- in the model. This is a significant further step on the path to simulating lattice gauge theories in more than one spatial dimension, in cases where they cannot be efficiently simulated using standard Monte Carlo approaches due to the sign problem. Our results agree with exact results where they can be compared, and show reasonable behavior for larger system sizes where they cannot.

Our ansatz is a superposition of Gaussian states coupled to gauge configurations, which allows us to use the covariance matrix formalism in order to efficiently calculate observables. 
Since it is a superposition involving Gaussian states, our final ansatz $\ket{\psi}$ is not itself Gaussian. This has proven sufficient for generating the numerical results demonstrated here, including the interaction between gauge fields and matter; more complex LGTs may require extending the state search over superpositions of these $\ket{\psi}$, each with different parameters in $\mathcal{T}$, as discussed in~\cite{roose2025gaugingsuperposition}.
Together with Monte Carlo integration over the gauge configurations, our approach allows for a variational ground state search which demonstrates the ability of the ansatz to capture the low-energy physics of a $\mathbb{Z}_2$ lattice gauge theory with matter.
In total, this method achieves our three aims: efficient representation of and computation with the state, guaranteed gauge invariance, and the entanglement area law expected for the system under consideration.

The next step in applying GGPEPS to LGTs is their application to a model that suffers from the sign problem, such as one with several flavors of fermions with different chemical potentials. 
Recent updates to our simulation code have significantly improved its efficiency, which will allow for increasing the size and complexity of the systems that can be studied using this approach, as well as greater investigation of the numerical performance of the algorithms as a function of ansatz settings.
This will pave the way for studying theories with other gauge groups, culminating with the study of quantum chromodynamics in 3+1d.

\section{Data Availability}
\label{sec:data_availability}

All data presented above and in the appendices is available in a Zenodo repository at~\cite{zenodo_data}.
The repository also includes a script that can generate all the plots, and a helper script to manually explore the data.

\section{Code Availability}
\label{sec:code_availability}
Code is available upon request.

\section{Author Contributions}
\label{sec:contributions}
A.K. and P.E. developed the simulation code. A.K. ran the data collection and wrote the majority of the text. U.B. wrote the code for the ED results. E.Z. supervised the work and developed large parts of the theory. All authors revised the manuscript.

\section{Acknowledgments}
\label{sec:acknowledgments}

We would like to thank G. Roose, I. Gomelski, and J. Elyovich for fruitful discussions.

P.E. acknowledges the support received by the Dutch National Growth Fund
(NGF), as part of the Quantum Delta NL program. 
P.E. acknowledges further the support received through the NWO-Quantum Technology program (Grant No.~NGF.1623.23.006) and funding by the Carl-Zeiss-Stiftung (CZS Center QPhoton). 
U.B. acknowledges support from the Israel Academy of Sciences and Humanities through the Excellence Fellowship for International Postdoctoral Researchers. U.B. acknowledges funding by the Max Planck Society, the Deutsche Forschungsgemeinschaft under Germany’s Excellence Strategy – EXC-2111 – 390814868, and the European Research Council (ERC) under the European Union’s Horizon Europe research and innovation program (Grant Agreement No.~101165667)—ERC Starting Grant QuSiGauge.

E.Z. acknowledges the funding by the European Union (ERC, OverSign, 101122583). Views and opinions expressed are those of the authors only and do not necessarily reflect those of the European Union or the European Research Council.

The results were computed using computing resources at the Fritz Haber Center for Molecular Dynamics at Hebrew University, as well as the Academic Leiden Interdisciplinary Cluster Environment (ALICE) provided by Leiden University.

\section{Competing Interests}
The authors declare no competing interests.

\bibliography{references.bib}

\appendix
\renewcommand{\appendixname}{Supplementary Information}
\renewcommand{\figurename}{Supplementary Figure}
\renewcommand{\tablename}{Supplementary Table}
\setcounter{figure}{0}
\setcounter{table}{0}

\section{Ansatz Details}
\label{sec:ansatz-details}

In the main text, we focused on the most salient points of the ansatz --- this section gives a more complete description and adds to the presentation in the main text.
We start with a discussion of the number of virtual modes used in the construction.
It was mentioned in the~\nameref{sec:ansatz} section (\ref{sec:ansatz}) that the particular ansatz utilized here uses four copies of virtual modes per site per link. It was shown in~\cite{emonts_finding_2023} that for a pure gauge $\mathbb{Z}_2$ theory, in order for the ansatz to be expressive enough to capture the ground state limits of the (pure gauge) Hamiltonian, at least two copies are needed.
For the ansatz which includes physical matter, it would \textit{prima facie} be unnecessary to add in additional virtual modes. However, the inclusion of matter requires the consideration of a global $U(1)$ symmetry. To ensure that the constructed ansatz state is invariant under this symmetry, we require virtual modes that are charged under the symmetry. Thus, each virtual mode must fall into one of three categories~\cite{kelman_2024}:
\begin{enumerate}
    \item Those which are uncharged under the global $U(1)$ symmetry. 
    \item Those which are charged under the global $U(1)$ symmetry in the same way as the matter.
    \item Those which are charged under the global $U(1)$ symmetry in a complementary way (i.e. conjugate) to the matter, and so pick up the opposite phase to the matter.
\end{enumerate}
Note that after the particle hole transformation, the $U(1)$ symmetry acts differently on physical modes on the even/odd sublattices, as seen in equation~\eqref{eq:u1-after-ph}. A given copy of the virtual modes will similarly transform with opposite phases under the $U(1)$ transformation on the even/odd sublattices.
As a result of the $U(1)$ symmetry, modes --- physical or virtual --- can only couple to modes which pick up the opposite phase.
In order to allow our ansatz to be at least as expressible as the one considered in~\cite{emonts_finding_2023} --- which allowed all-to-all coupling of the modes, since they were all of the first type --- we therefore leave two copies of virtual modes to play the same role, and leave them uncharged by the $U(1)$ symmetry. This ensures that the ansatz can capture the pure-gauge physics of the system.
To accommodate virtual modes which couple to physical matter, we must therefore add a third copy. This is insufficient, however, since one copy would not allow for virtual-virtual coupling of modes on the same site as is done in the operator $A(\site{x})$. We therefore include a copy of each of the latter two types, bringing the total to four.

Once the number of virtual modes is specified, it remains to be shown how they transform under the symmetries of the state, and the constraints this imposes on the construction of the ansatz.
The modes transform under the $U(1)$ symmetry as just described --- copies 1 and 2 are uncharged, copy 3 transforms with the opposite phase of the physical modes, while copy 4 transforms in the same way as the physical modes. Thus copies 1 and 2 can couple freely to each other, but not to copies 3, 4, or the physical modes. Copy 3 can only couple to copy 4 modes belonging to the same site as well as the physical mode there, or --- since neighboring sites are never on the same even/odd sublattice --- to copy 3 modes of neighboring sites, as happens in the projector operators.

Therefore, the structure of $\mathcal{T}$, which couples the various modes in the operator $A$, is 
\begin{equation} 
\label{eq:T-structure}
    \mathcal{T} = \begin{pNiceArray}{c|cc:cc}[margin, first-row] 
        \psi & \text{copy 1} & \text{copy 2} & \text{copy 3} & \text{copy 4} \\
        0         & \mat{0}       & \mat{0}   & M     & \mat{0}  \\ \hline
        \mat{0}   & \Block{2-2}{V^{\text{PG}}} &           & \mat{0}  & \mat{0}  \\
        \mat{0}   &               &           & \mat{0}  & \mat{0}  \\ \hdottedline
        -M^\top   & \mat{0}  & \mat{0}  & \mat{0}   & \tilde{V}   \\ 
        \mat{0}   & \mat{0}  & \mat{0}  & - \tilde{V}^\top  & \mat{0}  \\ 
    \end{pNiceArray},
\end{equation}
where the blocks have further structure in order to guarantee the other symmetries, as discussed below.
Note that the gauging operators $\mathcal{U}$ and projectors $w$ only couple allowed modes, and so this completes accounting for the constraints arising from the $U(1)$ symmetry.
We have also taken into account the canonical fermionic anti-commutation relations.

In order to ensure translation invariance, it is sufficient to require that $\mathcal{T}(\site{x}) = \mathcal{T}$. This was already indicated by the suppression of site dependence in equation~\eqref{eq:T-structure}. Ensuring translation invariance does not impose any extra structure on the gauging or projector operators.

Finally, we consider invariance under rotations. We introduce the permutation matrix
\begin{equation}
\label{eq:rot-mat}
    \mathcal{R}_0 = \left( {\begin{array}{cccc}
    		0 & 1 & 0 & 0 \\
    		0 & 0 & 1 & 0 \\
    		0 & 0 & 0 & 1 \\
    		1 & 0 & 0 & 0 \\
    \end{array} } \right),
\end{equation}
which rotates a single copy of the virtual modes around a site (ordered right, up, left, down).
If we group the creation operators of all of the modes of a given site into a vector, the rotation is implemented by
\begin{equation} \label{eq:full-rot-mat}
    \mathcal{R} = \begin{pNiceArray}{c|ccc}[margin,] 
                \eta & & & \\ \hline
                & \eta_\mu R_0 & & \\
                & & \Ddots^{} & \\
                & & & \eta_{\mu'} R_0
            \end{pNiceArray},
\end{equation}
where there are four copies of $\mathcal{R}_0$ along the diagonal. Modes are ordered: physical, copy 1, 2, 3, 4, with the modes in each copy ordered: right, up, left, down. The phases $\eta_\mu$ are all taken to be $e^{i\pi/4}$, i.e. the same as $\eta$. This is a choice, but one that is consistent with the construction of the ansatz.

Inspection reveals that the gauging and projecting operators are invariant under such rotations (the operators switch which link they act on, but retain their form). It therefore remains to show that $A$ remains invariant, for which we require that 
\begin{equation} 
\label{eq:T-rot-condition}
    \mathcal{R^\top TR} = \mathcal{T}.
\end{equation} 
Imposing this condition gives structure to the submatrices in $\mathcal{T}$ shown in equation~\eqref{eq:T-structure}. The elements which couple the first two copies --- ``PG'' denotes ``pure gauge'' to indicate that no physical matter is included --- have the form
\begingroup\makeatletter\def\f@size{9}\check@mathfonts
\begin{equation}
  V^{\text{PG}} = \begin{pNiceArray}{cccccccc}[margin, first-row]
           r_1   & u_1   & l_1    & d_1   & r_2   & u_2  & l_2    & d_2   \\ 
           0     & -z_1  & -i y_1 & -i z_1& i a   & i b  & i c    & i d   \\
           z_1   & 0     & -i z_1 & y_1   & -d    & -a   & -b     & -c    \\
            i y_1 & i z_1 & 0      & z_1   & -i c  & -i d & -i a   & -i b  \\
           i z_1 & -y_1  & -z_1   & 0     & b     & c    & d      & a     \\
           -i a  & d     & i c    & -b    & 0     & -z_2  & -i y_2  & -i z_2 \\
           -i b  & a     & i d    & -c    & z_2   & 0    & -i z_2  & y_2    \\
           -i c  & b     & i a    & -d    & i y_2 & i z_2 & 0      & z_2    \\
           -i d  & c     & i b    & -a    & i z_2 & -y_2  & -z_2    & 0     \\
          \end{pNiceArray},
\end{equation}
\endgroup
while the copies which do couple to physical matter are given by
\begin{equation} \label{eq:Tmat-rot-symm-offdiag}
    \tilde{V} = \begin{pNiceArray}{cccc}[margin]
                if  & ig  & ih  & ik \\
                -k  & -f  & -g  & -h \\
                -ih & -ik & -if & -ig \\
                g   & h   & k   & f \\
            \end{pNiceArray},
\end{equation}
and finally,
\begin{equation}
    M = 
    \begin{pNiceArray}{cccc}[margin]
        it & -t  & -it & t \\
    \end{pNiceArray}.
\end{equation}
All of the parameters are complex-valued.

The projectors defined in equations~\eqref{eq:projector-operators-layer1}, \eqref{eq:projector-operators-layer2}, and \eqref{eq:projector-decomposition} also satisfy all of the symmetries just mentioned. 
Note that the projectors are here defined as the Hermitian conjugate of those given in~\cite{emonts_finding_2023}.

Since modes of copies 1-2 never couple to modes of copies 3-4, the state can be factored into $\psi_I(\mathcal{G})$ and $\ket{\psi_{II}(\mathcal{G})}$ as described above. All of the calculations, including of covariance matrices, can therefore be done independently, which lowers the maximum dimension of the matrices that must be dealt with~\cite{kelman_2024}.

\section{Observables}
\label{sec:observables}

In this section we show how to calculate observables for a fixed gauge field configuration from covariance matrices, which can be found as described in~\cite{kelman_2024}. It was shown in~\cite{emonts_finding_2023} how to evaluate the operators on the gauge fields, including the magnetic and electric energies as well as their gradients. The calculation of the electric energy is described using the projectors of equation~\eqref{eq:projector-operators-layer1}.
To adapt the calculation to account for the different projectors used in the construction of $\ket{\psi_{II}(\mathcal{G})}$, one can follow the same procedure, with the indices appropriate to the modes used in the definition of the projectors in equation~\eqref{eq:projector-operators-layer2}.
Here we demonstrate how to evaluate terms involving matter. 
All results in this section are after the particle hole transformation of the~\nameref{sec:particle-hole} section. 

We start by defining Majorana modes, as the entries in the resulting covariance matrix are purely real. They are defined in terms of the Dirac modes as
\begin{equation} \begin{aligned}
\label{eq:dirac-majorana-modes}
    c^\dagger &= \frac{1}{2} (\gamma^{(1)} + i\gamma^{(2)})
    \qquad \quad &
    \gamma^{(1)} &= c + c^\dagger \\
    c &= \frac{1}{2} (\gamma^{(1)} - i\gamma^{(2)})
    \qquad \quad &
    \gamma^{(2)} &= i(c - c^\dagger) \\
\end{aligned} \end{equation}
where $c^\dagger, c$ are Dirac creation and annihilation operators, and $\gamma^{(1)}, \gamma^{(2)}$ are the corresponding Majorana modes.

The covariance matrix of a state $\ket{\psi(\mathcal{G})}$ is then defined as
\begin{equation} \begin{aligned}
\label{eq:physical-fermions-cov}
    \Gamma_{\site{x}^i, \site{y}^j}(\mathcal{G})
        &= \frac{i}{2} \langle [\gamma^{(i)}(\site{x}), \gamma^{(j)}(\site{y})] \rangle \\
        &= \frac{i}{2} \frac{\bra{\psi(\mathcal{G})}[\gamma^{(i)}(\site{x}), \gamma^{(j)}(\site{y})] \ket{\psi(\mathcal{G})}}{\braket{\psi(\mathcal{G})}}, \\
\end{aligned} \end{equation}
and, if the state is Gaussian (as our ansatz is, for a fixed~$\mathcal{G}$), this contains all the information required to reconstruct the state.

It remains to show how to calculate the mass and interaction terms of the Hamiltonian.
We start by rewriting the expectation value of the mass term of the Hamiltonian as
\begin{equation} \begin{aligned}
\label{eq:mass-term-simplified}
    \langle H_\textnormal{M} \rangle
        &= \frac{\bra{\psi} H_\textnormal{M} \ket{\psi}}{\braket{\psi }} \\
        &= \sum_{\mathcal{G}, \mathcal{G}'} \frac{ \bra{\mathcal{G}'} \bra{\psi_{II}(\mathcal{G}')} \psi_I^*(\mathcal{G}') H_\textnormal{M} \psi_I(\mathcal{G}) \ket{\psi_{II}(\mathcal{G})} \ket{\mathcal{G}} }{\braket{\psi}} \\
        &= \sum_\mathcal{G} \frac{ \bra{\psi_{II}(\mathcal{G})} H_\textnormal{M} \ket{\psi_{II}(\mathcal{G})} } {\braket{\psi_{II}(\mathcal{G})} } p(\mathcal{G}) \\
\end{aligned} \end{equation}
using the probability defined in equation~\eqref{eq:probability} and where $\ket{\psi(Q)} = \psi_I(Q) \ket{\psi_{II}(Q)}$.
This is only possible because $H_\text{M}$ only acts on the Hilbert space of $\ket{\psi_{II} }$.
We define
\begin{equation} \begin{aligned}
\label{eq:F-M}
    \mathcal{F}_\text{M}(\mathcal{G})
        &= \frac{ \bra{\psi_{II}(\mathcal{G})} H_\textnormal{M} \ket{\psi_{II}(\mathcal{G})} } {\braket{\psi_{II}(\mathcal{G})} } \\
\end{aligned} \end{equation}
which can be written in terms of the Majorana covariance matrix as
\begin{equation} \begin{aligned}
\label{eq:mass-term-cov}
    \mathcal{F}_\text{M}(\mathcal{G})
        &= \frac{1}{2} \sum_\site{x} \Big( 1 + \Gamma_{\site{x}^2 \site{x}^1}(\mathcal{G}) \Big)
\end{aligned} \end{equation}
where $\site{x}^i$ indicates the index corresponding to the $i^\text{th}$ Majorana mode on site $\site{x}$.
To derive this result, rewrite $H_\text{M}$ as 
\begin{equation}
    H_\text{M} = \sum_\site{x} \frac{1}{2} \Big( 1 + [\psi^\dagger(\site{x}), \psi(\site{x})] \Big)
\end{equation}
using the anticommutation relation $\{ \psi^\dagger(\site{x}), \psi(\site{x})\} = 1$, and then write $\psi^\dagger(\site{x}), \psi(\site{x})$ in terms of Majorana modes. The result gives the elements of the covariance matrix as in equation~\eqref{eq:mass-term-cov}.
The full mass energy is thus given by
\begin{equation} \begin{aligned}
    \langle H_\text{M} \rangle
        &= \sum_Q \Bigg[ \frac{1}{2} \sum_\site{x} \Big( 1 + \Gamma_{\site{x}^2 \site{x}^1}(\mathcal{G}) \Big) \Bigg] p(\mathcal{G}). \\
\end{aligned} \end{equation}

A similar procedure allows for the evaluation of the interaction energy.
The interaction energy can be put in the form of equation~\eqref{eq:obs-expectation} as was done in the case of the mass energy (equations~\eqref{eq:mass-term-simplified} and~\eqref{eq:F-M}), since
\begin{equation} \begin{aligned}
    \langle H_{\text{I}} \rangle
        &= \frac{\bra{\psi} H_\text{I} \ket{\psi}}{\braket{\psi }} \\
        &= \sum_{\mathcal{G}, \mathcal{G}'} 
            \frac{ \bra{\mathcal{G}'} \bra{\psi_{II}(\mathcal{G}')} \psi_I^*(\mathcal{G}') H_\text{I} \psi_I(\mathcal{G}) \ket{\psi_{II}(\mathcal{G})} \ket{\mathcal{G}}}{\braket{\psi}} \\
        &= \sum_\mathcal{G} \frac{ \abs{\psi_I(\mathcal{G})}^2 \bra{\mathcal{G}} \bra{\psi_{II}(\mathcal{\mathcal{G}})} H_\text{I} \ket{\psi_{II}(\mathcal{G})} \ket{\mathcal{G}} }{\braket{\psi}} \\
        &= \sum_\mathcal{G} \frac{\bra{\mathcal{G}} \bra{\psi_{II}(\mathcal{G})} H_\text{I} \ket{\psi_{II}(\mathcal{G})} \ket{\mathcal{G}} }{ \braket{\psi_{II}(\mathcal{G})} } p(\mathcal{\mathcal{G}}), \\
\end{aligned} \end{equation}
so we define
\begin{equation} \begin{aligned}
\label{eq:F-I}
    \mathcal{F}_\text{I}(\mathcal{G})
        &= \frac{\bra{\mathcal{G}} \bra{\psi_{II}(\mathcal{G})} H_\text{I} \ket{\psi_{II}(\mathcal{G})} \ket{\mathcal{G}} }{ \braket{\psi_{II}(\mathcal{G})} }. \\
\end{aligned} \end{equation}
To evaluate this expression, note that for a given gauge field configuration $\mathcal{G}$, in the magnetic basis
\begin{equation} \begin{aligned}
    \langle {U(\site{x}, k)} \rangle
        &= e^{i \delta q(\site{x}, 1)}
\end{aligned} \end{equation}
where $\delta = \pi$, $q(\site{x}, 1)$ is the value that $\mathcal{G}$ assigns to the link $\ell = (\site{x}, k)$, and 
\begin{equation} \begin{aligned}
    \langle \psi^\dagger(\site{x}) \psi^\dagger(\site{y}) \rangle
        &= \frac{1}{4i} \Big( \Gamma_{\site{x}^1 \site{y}^1} + i\Gamma_{\site{x}^1 \site{y}^2} + i\Gamma_{\site{x}^2 \site{y}^1} 
            - \Gamma_{\site{x}^2 \site{y}^2} \Big)_\mathcal{G} \\
    \langle \psi(\site{x}) \psi(\site{y}) \rangle
        &= \frac{1}{4i} \Big( \Gamma_{\site{x}^1 \site{y}^1} - i\Gamma_{\site{x}^1 \site{y}^2} - i\Gamma_{\site{x}^2 \site{y}^1} - \Gamma_{\site{x}^2 \site{y}^2} \Big)_\mathcal{G}
\end{aligned} \end{equation}
using the same procedure involving commutators and Majorana modes used to evaluate the mass term. When this expression is used below, the expectation will be taken relative to the state $\ket{\psi_{II}(\mathcal{G})}$, and so all covariance matrices depend on $\mathcal{G}$, though this was suppressed to declutter the notation ($\psi_I(\mathcal{G})$ does not contribute, as can be seen from its absence in equation~\eqref{eq:F-I}, and as expected due the to absence of fermionic matter in $\psi_I(\mathcal{G})$).

Thus, the expectation of the interaction energy is given by
\begin{equation} \begin{aligned}
    \langle H_\text{I} \rangle
        &= \sum_\mathcal{G} \Bigg[ 
            \sum_{\site{x}} \frac{1}{2} \Big( 
            e^{i\pi q(\site{x}, 1)} \big[ \Gamma_{\site{x}^1, (\site{x}+\uvec{e}_1)^1} - \Gamma_{\site{x}^2, (\site{x}+\uvec{e}_1)^2} \big]
            \\ & 
            - e^{i\pi q(\site{x}, 2)} \big[ \Gamma_{\site{x}^1, (\site{x}+\uvec{e}_2)^2} + \Gamma_{\site{x}^2, (\site{x}+\uvec{e}_2)^1} \big] \Big)
            \Bigg]_\mathcal{G} p(\mathcal{G})
\end{aligned} \end{equation}
where the large bracketed term is $\mathcal{F}_\text{I}(\mathcal{G})$.

Note that as explained in~\cite{kelman_2024}, the action of any observable can be separated into its action on the various layers of the ansatz --- in our case $\psi_I(\mathcal{G})$ and $\ket{\psi_{II}(\mathcal{G})}$. 

To calculate the derivatives of the mass and interaction energies with respect to the parameters of the ansatz, it is sufficient to find the derivative of the covariance matrix of the state, which can be done using the expression for the covariance matrix found in~\cite{kelman_2024}.

\section{Free Fermion Limit}
\label{sec:free-fermions}

We now consider the free fermions limit of our model, which obtains when only the interaction coupling of the Hamiltonian is nonzero. This provides a simple benchmark for the ansatz for large systems ($6 \times 6$ and larger). It further provides an indication of when finite size effects are significant in our model.

In the absence of electric coupling ($g_E=0$), the gauge fields enter the Hamiltonian \eqref{eq:hamiltonian} only through the operator $U=\sigma^x$, and can therefore be regarded as a static background. In this section we also set $g_M=g_B=0$, so that the fermions are massless and no flux configuration is \textit{a priori} preferred. The phases in the interaction (hopping) term of equation~\eqref{eq:interaction-energy} give rise to a $\pi$-flux per plaquette, which results in a Dirac-like band structure and guarantees the correct continuum limit from the high-energy physics perspective. While a background $\mathbb{Z}_2$ gauge field can in principle modify the flux by adding a $\pi$ phase to any of the hoppings, this is guaranteed not to happen by Lieb's theorem \cite{lieb1994flux}. This states that at half filling, the energetically preferred hopping configuration for free fermions on the square lattice is the one realizing a $\pi$-flux per plaquette. 

The Hamiltonian of equation~\eqref{eq:interaction-energy} is diagonal in momentum space, with a dispersion relation
\begin{equation}
    \label{eq:dispersion_dirac}
    E(k_x,k_y) = \pm 2 g_I \sqrt{\sin^2{k_x}+\sin^2{k_y}},
\end{equation}
which exhibits Dirac cones at $\vec{\mathbf{k}} =(0,0)$ as expected. In position space the system has a $2\times 2$ unit cell, leading to a reduced Brillouin zone. Both $k_x$ and $k_y$ are restricted to take values between $0$ and $\pi$, and the points $(k_x, k_y)$ and $(k_x+\pi, k_y+\pi)$ are identified with each other. On a torus of size $L_x=L_y=L$ the momentum in either direction is quantized in units of $2\pi/L$. Depending on whether periodic or anti-periodic boundary conditions are imposed on the fermionic wave-function, it can take the values
\begin{equation}
    k_n = \left\{ \frac{2\pi n}{L} \right\} \qquad \qquad \text{(PBC)}
\end{equation}
or
\begin{equation}
    k_n = \left\{ \frac{2\pi (n+1/2)}{L} \right\} \qquad \text{(ABC)}
\end{equation}
respectively, with $n \in \{0,1, \dots L-1 \}$. We need to account for both possibilities and see which one provides the lowest energy, corresponding to the true ground state of the system. This is obtained by filling the energy bands \eqref{eq:dispersion_dirac} with the available momentum states starting from the bottom, located at $k_x=k_y=\pi/2$, up to $E=0$ where the Dirac cones touch. The results are summarized in supplementary table \ref{tab:energies_free_fermions}, from which it results that the lowest energy always corresponds to anti-periodic boundary conditions. The ansatz described in the main text matches these predictions closely even in the computationally challenging case of a $6\times6$ system.

\begin{table}[h]
    \centering
    \vspace{5mm}
    \begin{tabular}{l||c|c|c}
    & Energy (PBC)  & Energy (ABC) & GGPEPS
    \\ \hline \hline 
    $L=2$      & $0$ & \makecell{ $-4\sqrt{2}$ \\ $\approx -5.6569$} & $-5.6569$
    \\ \hline
    $L=4$   & \makecell{$-4(2+\sqrt{2})$ \\ $\approx -13.6569$} & $-16$ & $-15.958$
    \\ \hline
    $L=6$ & \makecell{$-8 (\sqrt{3}+\sqrt{6})$ \\$\approx -33.4523$} & \makecell{$-4(3\sqrt{2} +2 \sqrt{5})$ \\ $\approx -34.8591$} & $-34.353$
    \\ \hline
    \end{tabular}
    \caption{Ground state energies of the Hamiltonian \eqref{eq:interaction-energy} on tori of size $L_x=L_y=L$ for periodic (left) and anti-periodic (right) boundary conditions for the fermions. The true ground state always corresponds to anti-periodic boundary conditions.}
     \label{tab:energies_free_fermions}
\end{table}

For larger systems, the ground state energy of the free fermions case is shown in supplementary figure~\ref{fig:free_fermions}.
In addition to providing a check on the GGPEPS numerical results for larger systems, this may also provide an indication for when finite-size effects are significant.
While it is possible that finite-size effects are stronger for observables other than the energy, two considerations alleviate this concern: (i) as shown in figure~\ref{fig:energy-observables}, our state accurately captures each term in the Hamiltonian, showing that our optimization captures the underlying physics and not solely the total energy; (ii) as shown in figure~\ref{fig:wilson-larger-systems}, even at the lattice sizes considered here, we are able to observe the behaviour of observables in different regimes.

\begin{figure}[h]
	\includegraphics[width=1.0\linewidth]{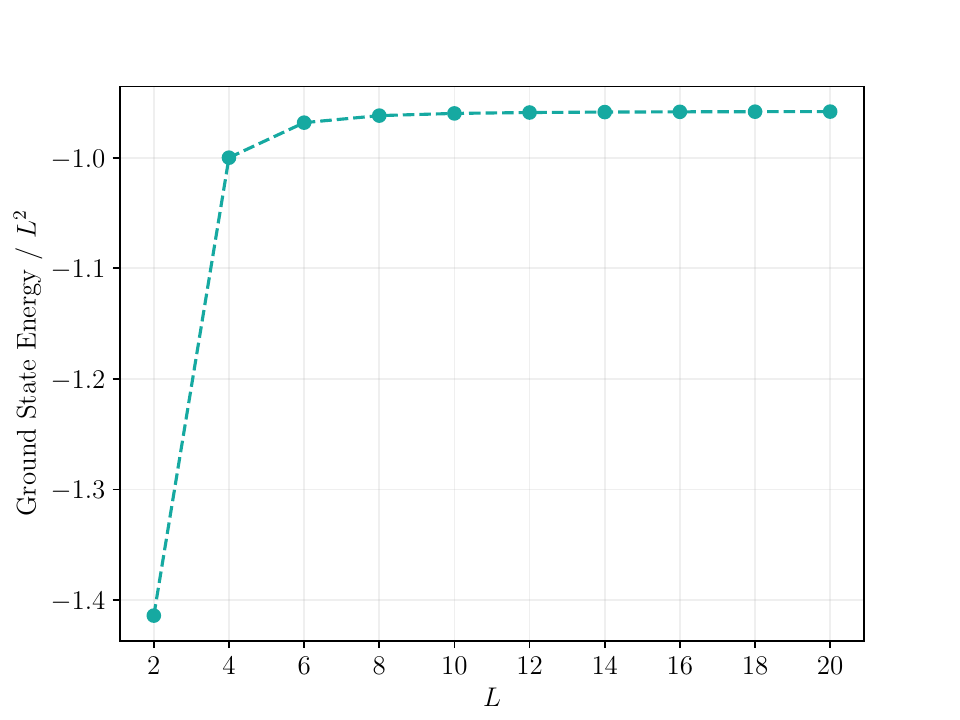}
	\caption{The ground state energies, normalized by lattice size, for the free fermion Hamiltonian. The dashed lines are simply guides for the eye. Even for moderate lattice sizes, the normalized ground state energy is relatively close to its value in the thermodynamic limit.}
    \label{fig:free_fermions}
\end{figure}

\section{Computation Details}
\label{sec:computation-details}

In this section, we provide some technical details on the configuration of our minimization scheme as well as Monte Carlo. Not all runs were run with identical settings, particularly as some points did not converge well on a first attempt, and were therefore rerun for further optimization. 

The minimization algorithm used was BFGS, using the standard implementation of \texttt{scipy}, available through \texttt{scipy.optimize.minimize} with the default settings~\cite{2020SciPy-NMeth}.

The number of Monte Carlo thermalization (warmup) steps was 50000 -- 100000 (starting with the latter figure, but moving to the former as experiments confirmed that this was sufficient). The number of measurement steps was generally 120000. At each Monte Carlo step, some number of links were chosen to have their gauge fields randomly modified. This update size was 8 for the $4\times4$ system, and $20$ for the $6\times6$ system.

\end{document}